\documentclass[twocolumn]{revtex4}
\usepackage{graphicx}

\makeatletter

\begin{document}

\title{Fluctuating epidemics on adaptive networks}

\author{Leah B.~Shaw}

\affiliation{Department of Applied Science, College of William and Mary,
Williamsburg, VA 23187}

\author{Ira B.~Schwartz}

\affiliation{US Naval Research Laboratory, Code 6792, Nonlinear Systems
  Dynamics Section, Plasma Physics Division, Washington, DC 20375}

\begin{abstract}
A model for epidemics on an adaptive network is considered.  Nodes follow an
SIRS (susceptible-infective-recovered-susceptible) pattern.  Connections are
rewired to break links from non-infected nodes to infected nodes and are reformed
to connect to other non-infected nodes, as the nodes that are not infected try
to avoid the infection.  Monte Carlo simulation and numerical solution of a
mean field model are employed.  The introduction of rewiring affects both the
network structure and the epidemic dynamics.  Degree distributions are
altered, and the average distance from a node to the nearest infective
increases.  The rewiring leads to regions of bistability where either an
endemic or a disease-free steady state can exist.  Fluctuations around the
endemic state and the lifetime of the endemic state are considered.  The
fluctuations are found to exhibit power law behavior.
\end{abstract}

\maketitle

\section{Introduction}

The study of recurrent epidemics has a long history \cite{Anderson91},
and many models, both deterministic and stochastic, have been considered.
Deterministic models have been used since the time of Bernoulli and
have explained some of the mechanisms in the spread of the infectious
diseases. However, deterministic models are not sufficient to account for some
of the important stochastic dynamics, such as extinction \cite{verdaska05,keeling04}
and sustained fluctuations \cite{solari01}. From the general theory
of finite Markov chains \cite{bartlett49}, it was shown that in
stochastic models the probability of extinction is equal to one in
the asymptotic time limit.  Numerical \cite{billings02,west97,citeulike:cumminhgs2005}
and analytic \cite{jacquez93} comparisons of stochastic and deterministic
models have been performed. The numerical results hold for very small
amplitude noise as well as real finite noise. Deterministic SIS or
SIRS models result in an equilibrium endemic presence
of infectives for an appropriate choice of parameters. It is clear
that stochastic effects may result in very different dynamics from
deterministic models, particularly when fluctuations and/or extinction
occur.

More recently, the study of fluctuations and the spread of simple
models of epidemics have been simulated on large networks \cite{Pastor-SatorrasV01,KupermanA01,moreno2002,MooreN00,HufnagelBG04}.
In almost all of these network models, the epidemic propagates on a fixed
network. The epidemic dynamics is typically studied as an SIS
or SIR model, in which the population is large and isolated. In addition
to the dynamics on such fixed architectures, controls based on vaccination
have been considered as well \cite{ZanetteK02,KosinskiA04}.  Several recent
models have considered epidemics on a network that changes structure dynamically
according to rules that do not depend on the nodes' epidemic status \cite{FeffermanN07,VolzM07}.

In contrast to the models of a static network or models with externally applied
changes in structure, a new class of models
based on endemic SIS populations on an \textit{adaptive} network has
been recently introduced \cite{GrossDB06}. Changes to the network structure
are made in response to the epidemic spread and in turn affect future
spreading of the epidemic.  Here, the new parameter
is one that describes the rewiring rate of the network, which is controlled
by the fraction of susceptible (S)-infective (I) links. The network alters
dynamically when there are contacts between S and I, and social pressures
(the desire to avoid illness) rewire the contacts to be instead between S and S. Infections are
reduced due to isolation, and a new phenomenon occurs:  for appropriate choices
of parameters, bistability
between the disease free equilibrium and endemic state has been
observed. This is in contrast to  static networks in a large population, where
there is typically only a
single attracting endemic or disease free state.

In this paper, we introduce a recovered, immune class and consider this slight
generalization of the SIS model on an adaptive network.  We examine the
structure of the network and the dynamics of the fluctuations of the epidemic.
Our approach is to combine Monte Carlo simulations and stochastic
mean field models for epidemic evolution on evolving networks. The
layout of the paper is as follows.  We introduce the model in Section
\ref{sec:model} and present its bifurcation structure in Section
\ref{sec:bifurcation}.  Properties of the network structure are discussed in
Section \ref{sec:network}.  We discuss dynamical properties of the system,
including fluctuations and lifetimes of the states, in Section \ref{sec:dynamics}.

\section{Model}

\label{sec:model}

We study an SIRS (susceptible-infective-recovered-susceptible) model on an
adaptive network.  Epidemic dynamics on the nodes is as follows.  The rate
for a susceptible node to become infected is $p N_{I,\text{nbr}}$, where
$N_{I,\text{nbr}}$ is the number of infected neighbors the node has.  The
recovery rate for an infected node is $r$.  A recovered node becomes
susceptible again with rate $q$, which we define as the re-susceptibility rate.  

While the epidemic spreads, the network is also being rewired adaptively.  If a link connects a non-infected node to an infected node, that link is rewired with rate $w$ to connect the non-infected node to another randomly selected non-infected node.  Self links and multiple links between nodes are disallowed.

In examining steady state solutions, it is sufficient to fix one of the rates, as time may be rescaled accordingly.  For this reason, we fix $r=0.002$ throughout this paper.

We performed Monte Carlo simulations of this model on a system with $N=10^4$
nodes and $K=10^5$ links.  (Larger system sizes with the same node to link ratio were also considered.  The major results of this paper do not depend strongly on system size.)  In each Monte Carlo step (MCS), we randomly select
$N$ nodes and $M$ links, where $M$ is the number of links that may potentially
rewire (susceptible-infected and recovered-infected links), and the links are
selected from the pool of links that may rewire.  Initial conditions are
constructed in one of two ways.  We either generate a random (Erd\"{o}s-R\'{e}nyi) graph of susceptibles and convert a fraction $f$ of them to
infectives, or we use the final state of a previous run as an initial
condition.  Transients are discarded and simulations run long enough that the initial conditions do not affect the results.

Following \cite{GrossDB06}, we also developed a corresponding mean field model for the system.  The mean field model tracks the dynamics of both nodes and links.  $P_A$ denotes the probability of a node to be in state $A$, where $A$ is either S (susceptible), I (infected), or R (recovered).  $P_{AB}$ denotes the probability that a randomly selected link connects a node in state $A$ to a node in state $B$. We obtain the following mean field equations for the evolution of the nodes:
\begin{eqnarray}
\label{eq:mf_node}
\dot{P}_S &=& q P_R -p \textstyle{\frac{K}{N}} P_{SI} \\
\dot{P}_I &=& p \textstyle{\frac{K}{N}} P_{SI}-r P_I \\
\dot{P}_R &=& r P_I -q P_R 
\end{eqnarray}
For example, in the first equation, recovereds are converted to susceptibles with rate $q$, and infection spreads with rate $p$ along each susceptible-infected link.  Rewiring does not appear directly in the node equations, since rewiring operates on links, but it affects the system implicitly through the number of susceptible-infected links ($K P_{SI}$).  We next write a system of mean field equations for the links.  To close the system, we follow \cite{GrossDB06} and make the assumption for three point terms that $P_{ABC} \approx P_{AB} P_{BC} / P_B$.  This assumption leads to the following system of equations for links:
\begin{eqnarray}
\dot{P}_{SS} &=& q P_{SR} +w \frac{P_S}{P_S+P_R} P_{SI} -2p \textstyle{\frac{K}{N}} \frac{P_{SS}P_{SI}}{P_S} \\
\dot{P}_{SI} &=& 2p\textstyle{\frac{K}{N}} \frac{P_{SS}P_{SI}}{P_S} +q P_{IR} -r P_{SI} -w P_{SI} \nonumber \\
&& -p \left( P_{SI} + \textstyle{\frac{K}{N}} \frac{ P_{SI}^2}{P_S} \right) \\
\dot{P}_{II} &=& p \left( P_{SI} + \textstyle{\frac{K}{N}} \frac{ P_{SI}^2}{P_S} \right) - 2r P_{II} \\
\dot{P}_{SR} &=& r P_{SI}+w \frac{P_R}{P_S+P_R} P_{SI} +2q P_{RR}-qP_{SR} \nonumber \\
&& -p \textstyle{\frac{K}{N}} \frac{P_{SI} P_{SR}}{P_S} +w \frac{P_S}{P_S+P_R} P_{IR} \\
\dot{P}_{IR} &=& 2r P_{II} +p \textstyle{\frac{K}{N}} \frac{P_{SI} P_{SR}}{P_S} -q P_{IR} -r P_{IR} \nonumber \\
&& -w P_{IR} \\
\dot{P}_{RR} &=& r P_{IR} -2q P_{RR} +w \frac{P_R}{P_S+P_R} P_{IR} 
\label{eq:mf_links}
\end{eqnarray}

The ordinary differential equations of the mean field model can be integrated easily with any well known numerical integration technique.
We choose initial
  conditions so that we are  near an endemic state. We also note
  that the model does support solutions with negative values, but these are
  unphysical, and so we ignore them.  In the case of stochastic simulations,
we have considered the effects of both internal fluctuations, modeled as
multiplicative noise, as well as external fluctuations, modeled as additive
noise. We use a fourth order Runge-Kutta solver for each of these cases to
generate stochastic stimulations of the mean field.  We also
  tracked  the steady states as a function of parameters using a continuation package \cite{auto}.

\section{Bifurcation structure}

\label{sec:bifurcation}

\begin{figure}[tbp]
\includegraphics[width=2.5in,keepaspectratio]{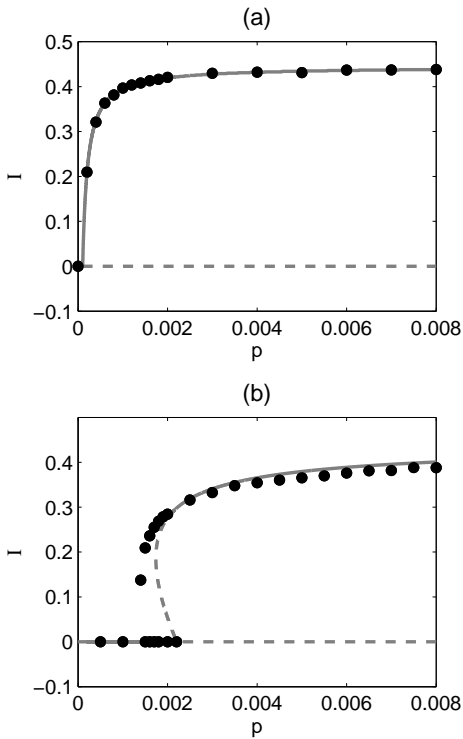}
\caption{Average infected fraction vs.~transmission rate $p$.  (a) Static
network, $w=0$.  (b) Rewired network, $w=0.04$.  Black dots:  Monte Carlo
simulations; solid gray lines:  mean field solution (stable branches); dashed
gray lines:  mean field solution (unstable branches). $q=0.0016$, $r=0.002$.}
\label{fig:bistable}
\end{figure}

We first consider the steady state bifurcation structure of the model.  In
Figure \ref{fig:bistable}, we show examples of the average infected fraction versus the
transmission rate $p$.  Two steady states can occur, a disease-free state and
an endemic state.  In the absence of rewiring (Fig.~\ref{fig:bistable}a), the
disease-free state loses stability for very small transmission rates, and only
the endemic state is observed at larger $p$ values.  When rewiring is
introduced (Fig.~\ref{fig:bistable}b), the disease-free state is stabilized for larger $p$ values.  A region of bistability, in which both endemic and
disease-free states are observed, now occurs.  Bistability was observed
previously for the SIS model \cite{GrossDB06}.  Monte Carlo simulation points in
Fig.~\ref{fig:bistable} were computed as follows.  To locate the upper
branches (endemic state), we swept $p$ from larger to smaller values, using
the final state of each run as the initial state for the next run.  $5 \times 10^3$ MCS of
transients were discarded, and the steady state was averaged over $10^4$ MCS.  To
locate the lower branch (disease free state), we began with a randomly
connected network in which a fraction $f$ of the nodes were infected, while the
rest were susceptible.  The system was simulated for $2 \times 10^4$ MCS.  $f$ values
between 0.025 and 0.9 were tried, and at least 5 runs were done for each $p$
value.  If the epidemic died out in any of the runs, the disease free state
was considered stable.  Due to the stochastic nature of the Monte Carlo
simulation, and because the disease-free state is absorbing, stability
designations are uncertain.  It is difficult to distinguish a weakly unstable
state from a weakly stable state with a short lifetime.  Lifetimes of the
endemic state are considered in more detail in Section \ref{sec:dynamics}.

The results in Figure \ref{fig:bistable} show fairly good agreement between the mean field approximation and the Monte Carlo
simulation of the full system, and we typically see this level of agreement in
the steady state values, although as we discuss later, the stability and type
of bifurcation sometimes differs.  Using the mean field model, we next explore
the bifurcation structure of the system for a wider range of parameters.

\begin{figure}[tbp]
\includegraphics[width=3.0in,keepaspectratio]{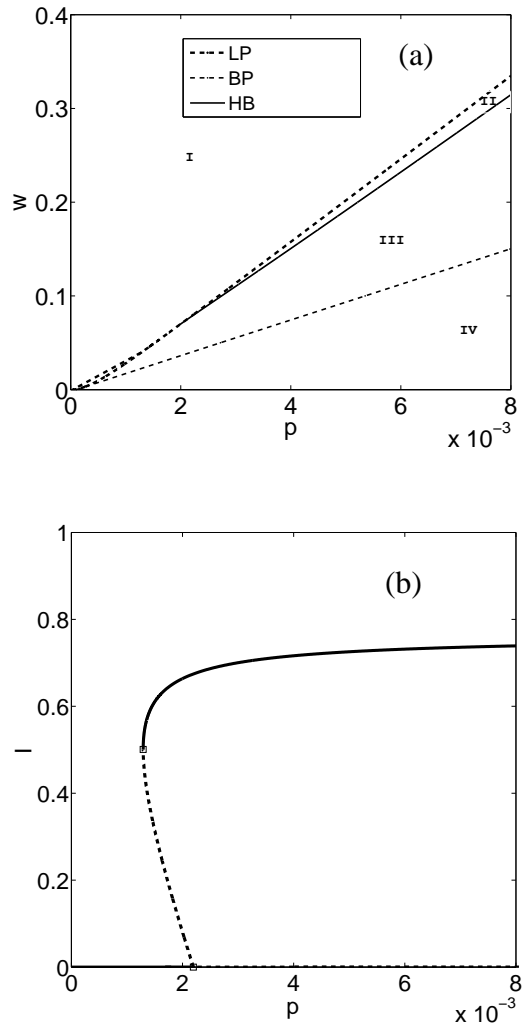}
\caption{(a) Two parameter plot for $w$ and $p$ of the bifurcation points for steady
  states when
   $q=0.0064$. The heavy dashed line is the line of saddle-node 
   points (limit points). The solid line denotes the Hopf bifurcation points, and
   light dashed line denotes the transcritical bifurcation points (BP). (b)  A
 a bifurcation diagram of the infective fraction  as a function of $p$, where $w =
 0.04$. The squares denote the saddle-node point and transcritical
point. Dashed lines are unstable branches.}
\label{fig:BDparamq0064}
\end{figure}

\begin{figure}[tbp]
\includegraphics[width=3.0in,keepaspectratio]{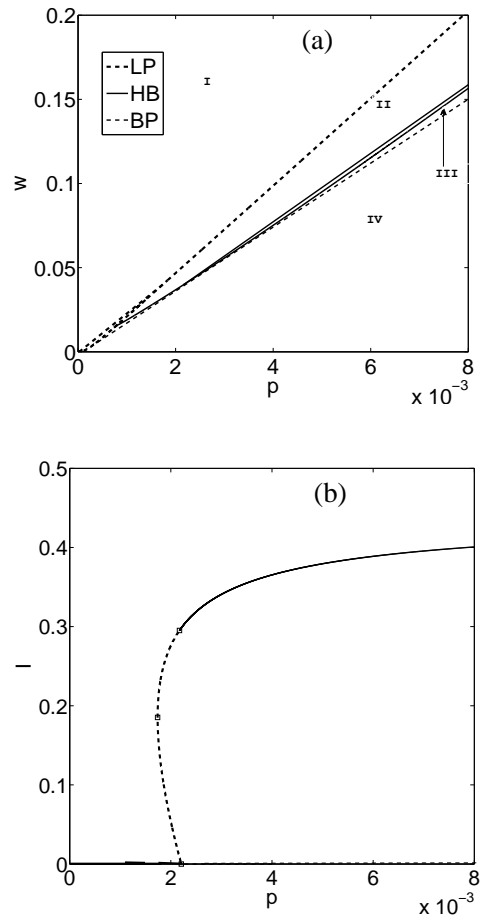}
\caption{(a) Two parameter plot of the steady state bifurcation points when
   $q=0.0016$. The heavy dashed line is the line of limit points.
 The solid line denotes the Hopf bifurcation points, and
   the light dashed line denotes the transcritical bifurcations. (b) A bifurcation diagram of the infective fraction  as a function of $p$, where $w =
 0.04$ (same bifurcation diagram in Fig.~\ref{fig:bistable}).}
\label{fig:BDparamq0016}
\end{figure}

An interesting property of the steady state instabilities appears when
one considers each of the steady state bifurcation points of the mean field
equations. (There do exist branches of periodic orbits, but since they occur
within a very small range of parameters, we ignore them in this paper. They
will be treated elsewhere.) If the re-susceptibility rate $q$ is held fixed
and a bifurcation diagram constructed, we find the existence of at least two
distinct regimes for different $q$ values,
illustrated in Figs.~\ref{fig:BDparamq0064} and \ref{fig:BDparamq0016}.
The   instabilities appear as a transcritical bifurcation from the
disease free state
steady state, a saddle-node bifurcation of endemic steady states, and a Hopf
bifurcation, from which  a branch of subcritical unstable periodic orbits
emanates (not shown).

In the case where $q=0.0064$, depicted in
  Fig.~\ref{fig:BDparamq0064}b for $w=0.04$, we show  a typical bifurcation plot of stable and unstable steady states
as the infection rate $p$ is increased. The dashed lines represent the
unstable steady states, while the solid lines depict the
attractors. For low
values of $p$ the disease free  steady state is stable. Tracking along the disease free branch, at a critical value of
$p$,  an unstable
branch (subcritical) of endemic steady states appears. The endemic branch
becomes stable 
 at the saddle-node point.
 There exists a clear region of bistability with
 coexisting endemic and disease free states for a range of $p$. If we now vary the parameter $w$ and track each bifurcation curve, we
obtain the result in Fig.~\ref{fig:BDparamq0064}a.  We describe the bifurcation
regions in detail for $w$ larger than 0.1.  In region I, we have only a stable disease free equilibrium. As we
cross into region II for large $w$, the disease
free equilibrium is
stable, and there exists an unstable endemic state. Region III exhibits
bistability between the disease free equilibrium and endemic state, and
region IV has just a stable endemic equilibrium. We note that at $w=0.04$, we
have the simple saddle-node transition depicted in Fig.~\ref{fig:BDparamq0064}b, since there is no Hopf bifurcation to periodic cycles at that particular $w$ value. 

The previous discussion presented a case for re-susceptibility rate $q$ where
the limit (saddle-node) and Hopf bifurcation branches are
close to each other. The distinction between the saddle-node and Hopf branches can be seen more
easily if $q$ is lowered to 0.0016, as shown in Fig.~\ref{fig:BDparamq0016}a. In Fig.~\ref{fig:BDparamq0016}a,
for sufficiently large $w$, as $p$ is increased in region I the system first undergoes a
limit point
bifurcation, and then
a Hopf bifurcation as it passes through region II. The Hopf curve is actually a closed isola in two parameters. The limit point here is a saddle-saddle
point, where a steady state having a  two dimensional unstable manifold connects to a
steady state with a one dimensional unstable manifold. In both cases, we have
bistable behavior for $w$ sufficiently large, but the region of bistability is
much smaller since the Hopf and transcritical branches are closer together for
this value of $q$ (region III).

For $w=0.04$, the mean field endemic steady state loses stability in a saddle-node
bifurcation for $q=0.0064$ and in a Hopf bifurcation for $q=0.0016$.  In our
discussion of fluctuations in the endemic state in Section \ref{sec:dynamics}, we
will refer to $q=0.0064$ because the saddle-node bifurcation structure best
corresponds to the scaling of fluctuations that we observe in Monte Carlo
simulations of the full system. 

\section{Network geometry}

\label{sec:network}

\subsection{Degree distributions}

\begin{figure}[tbp]
\includegraphics[width=3in,keepaspectratio]{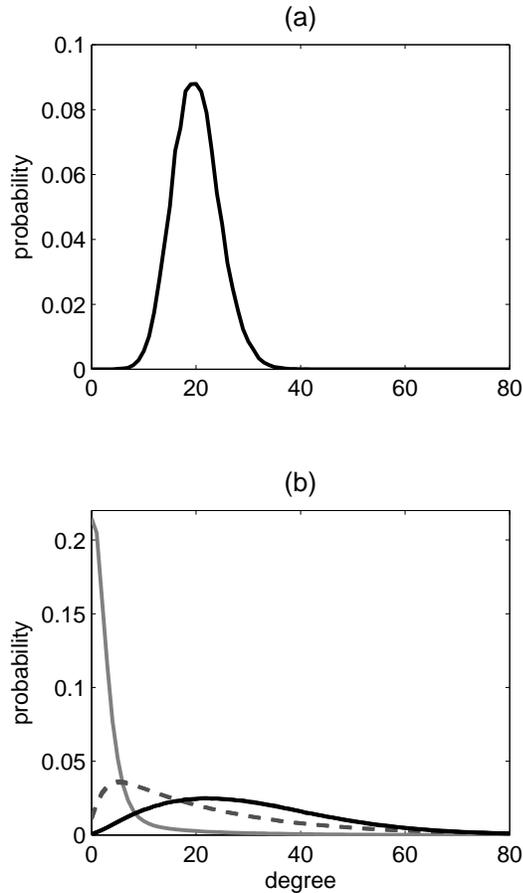}
\caption{Degree distributions from Monte Carlo simulation for $p=0.002$, $q=0.0016$, $r=0.002$.  (a) Static network, $w=0$.  All node types have Poissonian degree distribution.  (b) Rewired network, $w=0.04$.  Solid gray:  infectives; dashed:  recovereds; black:  susceptibles.}
\label{fig:degdistr}
\end{figure}

Rewiring leads to significant alterations in the network structure.  We first
consider the degree distribution.  Figure \ref{fig:degdistr} shows degree
distributions for each type of node in the absence (\ref{fig:degdistr}a) and
presence (\ref{fig:degdistr}b) of rewiring.  In Fig.~\ref{fig:degdistr}b, we
 averaged over $3 \times 10^4$ MCS.  In Fig.~\ref{fig:degdistr}a, the network is static, so we averaged over 10 separate runs to obtain better statistics.

Using mean field ideas, we can understand the recovered and susceptible degree distributions as nodes flow from infected to recovered to susceptible.  We outline the calculation briefly here.  However, this approach does not accurately predict the degree distribution for infected nodes because correlations play a more important role for these nodes, as we will explain below.  For this reason, we cannot write a self-consistent set of equations for the degree distributions that could be solved without inputting simulation results.

Let $d_{X,n}$ be the number of nodes in state $X$ (either S, I, or R) with degree $n$.  Recovered nodes originate when infectives recover and are lost when they become susectible again.  The degree of a recovered node can only increase, as other susceptible and recovered nodes wire to connect to it.  This leads to the following equations for $d_{R,n}$:
\begin{eqnarray}
\frac{d}{dt} (d_{R,0}) &=& r d_{I,0} - q d_{R,0} -k d_{R,0} \nonumber \\
\frac{d}{dt} (d_{R,n}) &=& r d_{I,n} -q d_{R,n} -k d_{R,n}  \label{eq:Rdeg} \\
& & +k d_{R,n-1} \; \; \text{for } n>0 \nonumber 
\end{eqnarray}
where $k$ is the average rate for nodes to rewire to a given non-infected node.  $k$ is given by the ratio of the total rewiring rate to the number of potential target nodes:
\begin{equation}
k = \frac{w K(P_{SI}+P_{IR})}{N(P_S+P_R)}.
\end{equation}
Given the degree distribution of the infectives $d_{I,n}$ and the probabilities appearing in $k$, Eqs.~\ref{eq:Rdeg} can be solved for the degree distribution of the recovereds.

\begin{figure}[tbp]
\includegraphics[width=3in,keepaspectratio]{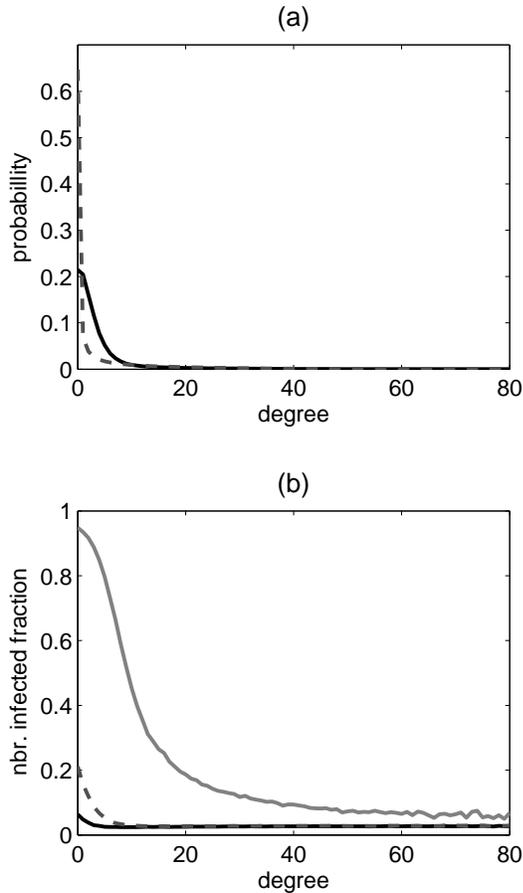}
\caption{(a) Actual (black) and predicted (dashed gray) degree distributions for infecteds. (b) Average fraction of neighbors that are infected vs.~degree.  Solid gray:  infecteds; dashed:  recovereds; black:  susceptibles.  $p=0.002$, $q=0.0016$, $r=0.002$, $w=0.04$. }
\label{fig:degcorr}
\end{figure}

Susceptible nodes originate when recovereds become susceptible again, and they may be lost when they become infected by a neighbor.  As with the recovereds, the degree of a susceptible increases due to rewiring.  Thus the time evolution of the degree distribution for the susceptibles can be written as
\begin{eqnarray}
\frac{d}{dt} (d_{S,0}) &=& q d_{R,0}  -k d_{S,0} \nonumber \\
\frac{d}{dt} (d_{S,n}) &=& q d_{R,n} -k' n d_{S,n} -k d_{S,n} \label{eq:Sdeg} \\
& & 
 +k d_{S,n-1} \; \; \text{for } n>0 \nonumber
\end{eqnarray}
where $k'$ is the infection rate per link into a susceptible node.  We assume that $k'$ is independent of degree, which we know from simulations is approximately correct (cf.~Fig.~\ref{fig:degcorr}), and write
\begin{equation}
k' = p \frac{P_{SI}}{P_{SI}+P_{SR}+2 P_{SS}}
\end{equation}
where the fraction is the ratio of the number of links that can transmit infection to the total number of links into a susceptible.  As with the recovereds, the steady state degree distribution for susceptibles can be computed from Eq.~\ref{eq:Sdeg}.  The predicted degree distributions for susceptibles and recovereds are overlaid on the actual distributions in Fig.~\ref{fig:degdistr}b, using Monte Carlo simulation averages for the infective degree distribution and the node and link probabilities.  (Note that the node and link probabilities could instead be obtained from the mean field system.)  Deviations between the prediction and simulation are smaller than the width of the curves in Fig.~\ref{fig:degdistr}b, so they are indistinguishable.

The degree of infected nodes, however, cannot be predicted by this approximate procedure.  We might expect that 
\begin{eqnarray}
\frac{d}{dt} (d_{I,0}) &=& k'' d_{I,1}  -r d_{I,0} \nonumber \\
\frac{d}{dt} (d_{I,n}) &=& k'' (n+1) d_{I,n+1} -k'' n d_{I,n} \label{eq:Ideg} \\
& &+k' n d_{S,n}
 -r d_{I,n} \; \text{for } n>0  \nonumber
\end{eqnarray}
where 
\begin{equation}
k'' = w \frac{P_{SI}+P_{IR}}{P_{SI}+P_{IR}+2 P_{II}} \label{eq:krewiring}
\end{equation}
is the per link rewiring rate for links connecting to an infective (i.e., the
ratio between links that can potentially rewire and total links into
infectives).  Figure \ref{fig:degcorr}a compares the actual degree
distribution for infectives with the distribution predicted by
Eq.~\ref{eq:Ideg}.  The number of low degree infectives is significantly
over-predicted.  This occurs because the mean field approximation in
Eq.~\ref{eq:krewiring} is not accurate for infectives.  Figure
\ref{fig:degcorr}b shows the fraction of infected neighbors that a node has,
depending on its degree and disease status.  Results are averaged over $3
\times 10^4$ MCS.  Low degree infected nodes tend to have a much higher
fraction of neighbors that are also infected, due to transmission of the
disease.  Once infected, these neighbors will not rewire away until recovered, so the rewiring rate per link is smaller than one would expect from the mean field $k''$ in Eq.~\ref{eq:krewiring}.  Predicting the infective degree distribution accurately would require a theory that accounts for these correlations in infection status of neighboring nodes, which is beyond the scope of the present work.

\subsection{Distance from an infective}

\begin{figure}[tbp]
\includegraphics[width=3in,keepaspectratio]{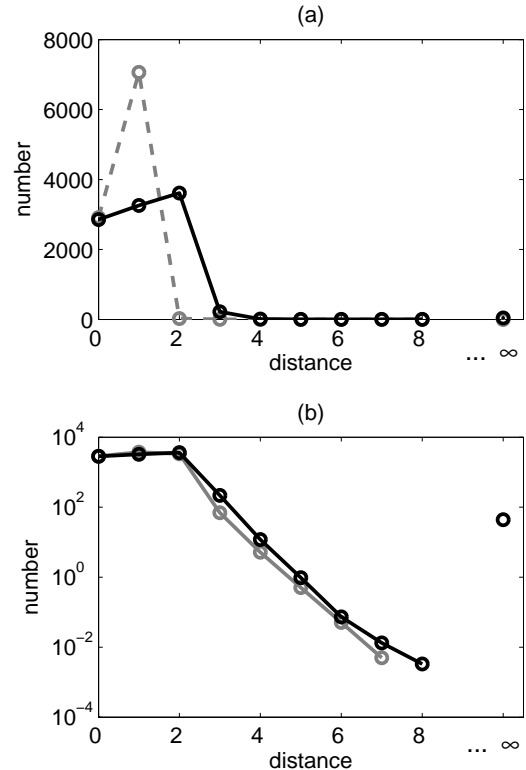}
\caption{Distribution of distances from the nearest infected node.  $\infty$ indicates nodes that are completely disconnected from an infective.  (a) Solid black:  with rewiring ($w=0.04$, $p=0.002$, $q=0.0016$); dashed gray:  no rewiring ($w=0$, $p=0.002$, $q=0.009$).  (b) Black:  rewired case (same as in (a)); gray:  distribution for random graphs, as described in text. }
\label{fig:idist}
\end{figure}

We next consider the distribution of distances from a given node to the
nearest infective.  These distances are of interest because they relate to the
number of hops the disease must make in order to reach an uninfected
individual.  The disease cannot propagate through recovered nodes until they
become susceptible again, so the distance from the nearest infective does not
necessarily correspond to a path for disease propagation.  However, we note
that rewiring acts only on links to infectives, and thus the chains of
susceptible and recovered links that this metric identifies will persist until
the infection leads to their interruption.  Figure \ref{fig:idist}a shows the
distribution of distances from the nearest infective in the presence and
absence of rewiring.  To display the effect on the network geometry alone,
rather than on the steady state number of infections as well, we have used a
smaller $q$ value for the $w=0$ case so that the total number of infectives is
approximately the same in both curves.  Results were averaged over $3 \times 10^4$ MCS, sampled every 100 MCS after removing transients.  Rewiring significantly decreases the number of nodes that are directly connected to an infective.  However, despite the rewiring, only a small fraction of nodes are fully disconnected from the infection.

Figure \ref{fig:idist}b shows a semilog plot of the $w=0.04$ case to display
the tail of the curve at larger distances.  An approximation based on random
networks is also shown.  Given how far from Poisson the degree distributions
are when the network is rewired, it is somewhat surprising that the form of
the decay in the distribution of distances can be predicted from random
networks.  Beginning with S, I, and R nodes in numbers matching that
observed in the average of Monte Carlo simulations, we generated 1000 random
networks and added randomly selected links until the number of S-S, S-I, S-R,
etc., links also matched that from the average of Monte Carlo simulations.  As
Figure \ref{fig:idist}b shows, the distribution of distances for these random
networks decays in the same way as it does in an adaptive network.  Thus the
form of the distribution of distances depends mainly on local dynamics (node
and link dynamics) rather than on the details of longer range correlations.
The main difference is that the adaptive network has some nodes fully
disconnected from infected components, while the random networks do not.  Most
of these nodes are recovereds with degree 0, which appear when infectives of
degree 0 recover.  Each type of node in the random networks has a Poisson degree
distribution, so they do not generally have nodes of degree 0.

\section{Fluctuations and other dynamics}
\label{sec:dynamics}

In the previous sections, we have considered steady states and long time averages of network properties.  We next consider fluctuations and dynamical properties of the endemic state.

\subsection{Fluctuations near bifurcation point}

\begin{figure*}[tbp]
\includegraphics[width=7in,keepaspectratio]{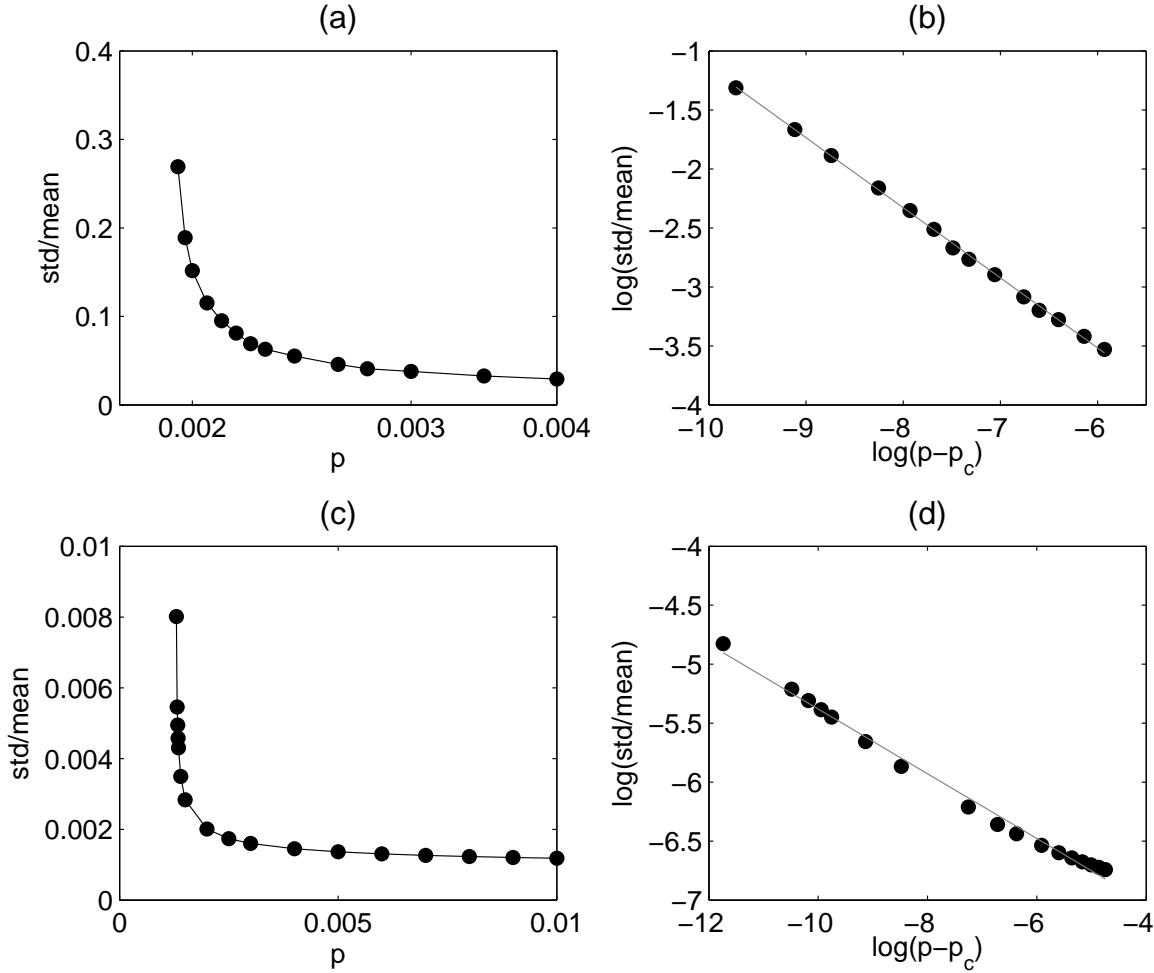}
\caption{Fluctuations in infectives (standard deviation divided by mean) vs.~infection rate $p$ near the bifurcation point:  (a) Monte Carlo, (b) mean field.  Curves are to guide the eye.  Log-log plots (data points with best fit lines) show power law scaling for both Monte Carlo (b) and mean field (d).  Monte Carlo parameters:  $q=0.0016, w=0.04, r=0.002$.  Mean field parameters:  $q=0.0064, w=0.04, r=0.002, \epsilon=0.0001$.}
\label{fig:fluct}
\end{figure*}

Near the bifurcation point where the endemic state loses stability, the number
of infectives has larger fluctuations due to noise overcoming weak
  attracting forces \cite{Horsthemke83}.  Fluctuations in the SIRS model are significantly larger than those in the previously studied SIS model.  We quantify the fluctuations by computing
the standard deviation divided by the mean for long time series in both the
Monte Carlo and mean field simulations.  In
Fig.~\ref{fig:fluct}, we plot the fluctuations as the infection rate $p$ is
swept towards the bifurcation point.  Monte Carlo results were computed from
$5 \times 10^5$ MCS time series sampled every 10 MCS, except for the two smallest $p$
values, for which shorter time series were used due to the shorter lifetimes
of these states.  All time series were longer than $10^5$ MCS.

For comparison, the mean field equations can also be considered near equilibrium in  stochastic 
form. In general, near equilibrium fluctuations can be modeled as additive
noise \cite{vanKampen_book}, and we do so here. (We do note that multiplicative
noise effects generate similar results to those reported for additive noise.) We assume the mean field is of the following form:
\begin{equation}
\bf{X}' = \bf{F(X)} + \epsilon \bf{\eta} (t),
\end{equation}
where $\bf{F(X)}$ is the mean field system in Eqs.~\ref{eq:mf_node}-\ref{eq:mf_links}, and  $\langle \bf{\eta} (t) \bf{\eta} (t') \rangle = \delta (t-t')$. $\epsilon$ is the noise
strength, or amplitude. 
We have considered both additive noise and multiplicative noise cases in 
the simulations of the stochastic attractors near the endemic state and have
computed the standard deviation divided by the mean as described above for 10 random initial conditions near equilibrium and 10 realizations.

We recall from  Figs.~\ref{fig:BDparamq0064} and \ref{fig:BDparamq0016} that depending on the value of
  the re-susceptibility rate $q$, the bistability regions III are vastly different. Specifically, for $q=0.0016$,
  we saw that for sufficiently large wiring rates, the saddle-saddle and Hopf
  bifurcation branches were well separated, whereas for $q=0.0064$, the
  branches were very close for small values of re-wiring rate $w$. In Monte Carlo simulation, we have not observed the Hopf bifurcations or stable periodic oscillations seen in the mean field, even for system sizes as large as $4 \times 10^5$ nodes.  Although
  the value of $q=0.0064$ we use in the mean field fluctuation study is different from that
  used in the Monte Carlo, the local bifurcation structure is similar when $w=0.04$,
  in that it is  a true saddle-node bifurcation point. This has been checked
  by examining the local linear vector field at the saddle-node point in question.
Therefore, although the mean field has different $q$ value, the bifurcation
structure is equivalent to that observed in Monte Carlo simulation, so we use $q=0.0064$ in
the fluctuation study.

The computations reveal that the 
fluctuations exhibit power law scaling, as shown in the log-log plots of
Fig.~\ref{fig:fluct}b and \ref{fig:fluct}d.  On the horizontal axis, we plot
$\ln (p-p_c)$, where $p_c$ is the critical point at which the endemic state
loses stability.  For the mean field, the bifurcation point is known exactly
by examining the eigenvalues of linearized vector field at the steady state.  We approximate the Monte Carlo bifurcation point $p_c$ as the value that
produces the most linear plot.  Although both cases have power law scaling,
the exponents are different:  $-0.59$ for Monte Carlo and $-0.27$ for mean field.  The scaling exponent for the full system depends on the number of nodes.  Whether it will approach the mean field value in the limit of infinite system size is a subject for further study.
We note that
fluctuations near a Hopf bifurcation point, as occurs in the mean field for
$q=0.0016$, would produce a very different form of scaling from the saddle-node case. One of the reasons for the difference is that the instability is
then two dimensional and underdamped \cite{arnold}. This is known to cause very different
scaling laws in generic problems, which can be much slower \cite{dykmanetal2005}.

To motivate the power law scaling of the fluctuations, we consider scaling
near a generic saddle-node bifurcation.  The simplest generic case of a saddle-node bifurcation for equilibrium
points comes from solving for zeroes of the vector field at a parameter
value where one eigenvalue of the Jacobian passes through zero. Standard normal form analysis
allows one to consider the generic problem of a saddle-node bifurcation.  In one dimension,
the stochastic differential equation of a saddle-node bifurcation
may be modeled as 
\begin{equation}
dx_{t}=(a-x_{t}^{2})dt
+\sigma*dW_{t}.
\label{Eq.SNSDE}
\end{equation}
The parameter is $a$, and we suppose noise is additive. Since
noise in general may cause a {}``shift'' in parameter values where
the saddle-node point disappears, we assume that the noise near the
bifucation is sufficiently small, where $dW/dt$ is a white noise
term, and $dW$ is a Brownian increment.

\begin{figure}
\includegraphics[width=3in]{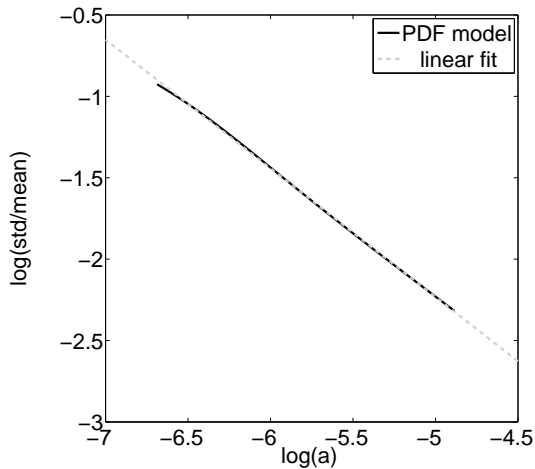}

\caption{\label{fig:SN_bifurcation_fluctuations} Fluctuation size of a generic
saddle-node bifurcation as a
function of bifuration parameter $a$ near the bifurcation point using Eq. \ref{Eq.SNPDF}.
The noise level used is 0.005.
}

\end{figure}

We further assume that we  are always near the attracting
branch of the saddle-node, so we are in a near equilibrium setting. Such an
assumption allows us to  examine the stationary probability
density function (PDF) of the stochastic dynamics by employing  the Fokker-Planck
equation near steady state. For the stochastic differential equation, Eq.~\ref{Eq.SNSDE}, the PDF is well known
\cite{Horsthemke83} and is given by 
\begin{equation}
p(a,x,\sigma)=Ne^{2(a x-x^{3}/3)/\sigma^{2}}.
\label{Eq.SNPDF}
\end{equation}
Here $N$ is a normalization constant. We compute the first and second
order moments directly using Eq.~\ref{Eq.SNPDF} and then take the ratio of the
standard deviation to the mean. Since $a=0$ is the value of
the saddle-node point, we examine the fluctuations in the neighborhood
of that value. The results, shown in
Fig. \ref{fig:SN_bifurcation_fluctuations}, display power law scaling.

\subsection{Delayed outbreaks}

\begin{figure}[tbp]
\includegraphics[width=3in,keepaspectratio]{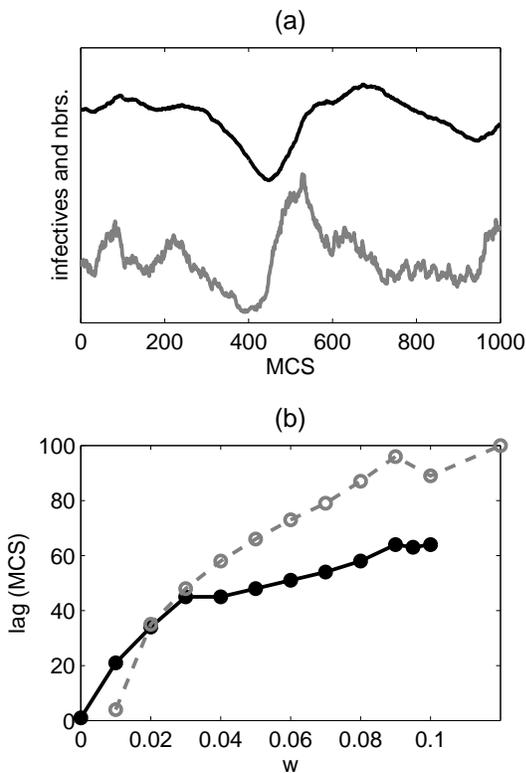}
\caption{Delayed outbreaks due to rewiring.  (a) Monte Carlo time series.  Black: infectives; gray:
neighbors of infectives.  Curves are scaled in arbitrary units for comparison
of peak times.  $p=0.0065, w=0.09, q=0.0016, r=0.002$.  (b) Time in MCS by
which infectives lag behind infective neighbors vs.~rewiring rate.  Solid
black:  Monte Carlo; dashed gray:  mean field.  $p=0.0065, q=0.0016,
r=0.002$, mean field noise strength $\epsilon=0.0001$. 
}
\label{fig:lags}
\end{figure}

We next consider phase relationships between the fluctuating variables.  We
tracked the number of infectives in the system at each time point as well as
the number of nodes that neighbored an infective (i.e., the number of SI and
IR links).  In the rewired system, fluctuations in the number of infectives
lagged behind fluctuations in the number of infective neighbors that are not
themselves infected, as shown in Fig.~\ref{fig:lags}a.

Both mean field and
Monte Carlo simulations of the full system displayed this effect, and we
studied its dependence on the rewiring rate.  Monte Carlo
simulations were sampled every 1 MCS for $3 \times 10^4$ MCS after discarding transients.
The mean field was sampled every 1 time unit for $5 \times 10^4$ units.
Additive noise was included in the mean field equations with noise strength
$\epsilon = 0.0001$.
Cross correlations between the infectives and the infective neighbors were computed for varying
shifts between the time series, and the lag maximizing the cross correlation
was identified.  As shown in Fig.~\ref{fig:lags}b, rewiring leads to
increasing lag times and delayed outbreaks.  We also computed time lags for the mean field model with multiplicative noise
and found the same trend of increasingly delayed outbreaks with larger
rewiring.  

\subsection{Lifetime of endemic steady state}

\begin{figure}[tbp]
\includegraphics[width=3in,keepaspectratio]{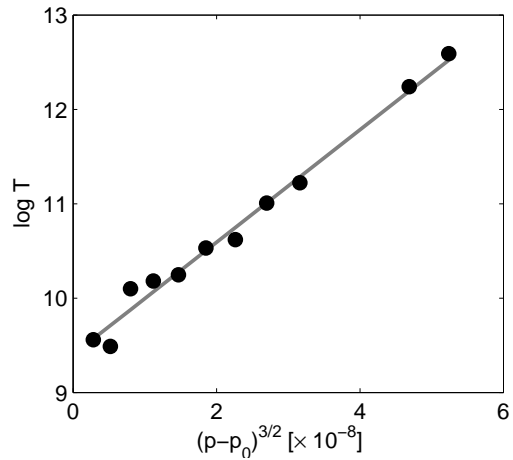}
\caption{Dependence of endemic state average lifetimes on infection rate
$p$. Points:  Monte Carlo simulations; line:  best fit line.  $q=0.0016$, $r=0.002$, $w=0.04$.  See text for details.  }
\label{fig:lifetimes}
\end{figure}

The final dynamic effect we consider is the lifetime of the endemic steady
state.  Because the system is stochastic and the disease free state is
absorbing, all parameter values will lead to eventual die out of the disease in the
infinite time limit.  These lifetimes become shorter and die out is more
easily observed in the bistable regime, near the bifurcation point where the
endemic state has weak stability.  We measured the dependence of the lifetimes on
the infection rate $p$.  For each $p$ value, we prepared a steady state
initial condition and computed multiple duplicate runs to obtain a distribution of
lifetimes.  (We computed 100 duplicate runs for all but the two highest $p$ values and over 25
runs for the two highest.)  We then calculated the average lifetime $T$ for each
$p$.  For a generic saddle-node bifurcation in one dimension, it is expected that 
 $\ln T$ is a linear function of $(p-p_0)^{3/2}$, where $p_0$ is the
bifurcation point \cite{Dykman1980,Graham1987a}.  We obtained spurious results for our system with $10^4$ nodes,
possibly because the small system size led to rapid die out and we were unable
to obtain good statistics near the bifurcation point.  Away from the
bifurcation point, the system has a weakly damped oscillatory component and behaves like a focus.  By switching to a
system with $4 \times 10^4$ nodes and $4 \times 10^5$ links, we were able to
run longer simulations closer to the bifurcation point and operate in a regime
where only one dimension mattered and the oscillations could be
ignored.
  Preliminary scaling results
are shown in Figure \ref{fig:lifetimes}.  We used the bifurcation point $p_0$
that gave the best fit line, which led to an $R$ value of 0.99.  The scaling
results appear consistent with expectations, but more study and better
statistics are needed.

\section{Conclusions and discussion}

\label{sec:conclusion}

We have explored the stable states, network properties, and dynamics of an
SIRS model on a network with adaptive rewiring.  As with the SIS model studied
previously by Gross \textit{et al.}~\cite{GrossDB06}, the rewiring leads to
bistability of the endemic and disease free states.  A mean field version of
the model predicted the steady states with good numerical accuracy and was
also valuable in studying the fluctuations of the system, with the caveat
that one must be near the appropriate type of bifurcation in the mean field to
obtain corresponding results.  With the addition of the recovered class and
  re-susceptibility rate, we can control the width of the bistability region
by manipulating the location of the bifurcation points.

  The fluctuations in the infectives near the
bifurcation point showed power law scaling.  This agreed with mean field
results and our expectations for scaling near a saddle-node bifurcation.

We studied the effects of the rewiring on the network geometry.  Degree
distributions were altered, and mean field arguments were able to predict the
distributions for susceptibles and recovereds.  However, a new analytical
approach that includes correlations is needed to fully understand the degree
distributions.

The other network property we considered was the distribution of distances from non-infected
nodes to the nearest infective, a quantity that may be important in disease
spreading and its control.  This distribution depended primarily on node and link dynamics
rather than on higher order correlations, so it
could be predicted from random graphs.  It is possible to compute the
distribution of distances from an infective analytically for random graphs,
but this calculation is awkward
for a three species system (S, I, R) and does not have a simple functional
form, so we have omitted discussion of analytical results here.

Delayed outbreaks were observed in the rewired system.  Peaks in the infective
fraction lagged behind peaks in the number of nodes that neighbor an infective.
For the parameter values studied here, the lag time is on the order of 10\% of
the mean infectious period, which might be considered a very short lag time.
However, the parameters in this study were not selected to correspond to any
specific disease.  For most real diseases, we would expect a much slower rate
for immunity to wear off and recovered individuals to become susceptible
again, compared to the mean infectious period.  This regime would be more
difficult to study in Monte Carlo simulations, since the average number of
infected nodes would be much smaller than seen here.  Further work is needed
to determine whether the observation of delayed outbreaks due to rewiring
would persist or perhaps become more significant in a physically realistic system.

Finally, we considered lifetimes of the endemic state near the saddle-node bifurcation
where it loses stability.  In order to achieve extinction from a steady state,
the disease must first overcome the attractive forces, which are weak near the
bifurcation point. Due to the generic local topology of the
saddle-node structure, the escape rate is well characterized analytically \cite{dykmanetal2005}. We
found that the Monte Carlo simulation agrees qualitatively with the escape times and yields a
well-known power law. However, this is for parameters in which endemic and
extinction states are not too far apart. Such extinction regimes can be analyzed using a
Fokker-Planck approach \cite{doeringetal2005}. Although preliminary results are in agreement with the
expected scaling,  further study of the lifetime scaling is needed, including
in regimes that are physically realistic, and where the usual extinction
rates cannot be modeled with a Fokker-Planck approach.

In addition to the directions for further research mentioned above, a major
challenge is to develop network geometries and rewiring rules that are more
consistent with human social networks.  Real social networks are expected to
have community structure \cite{GirvanN02}.   The networks studied here
did not.  In our rewired networks, the number of connections from non-infected
nodes to infected nodes was reduced in comparison to a random network, while
the number of connections between non-infected nodes was increased.  However,
this process does not induce a community structure on the network in the
Newman-Girvan Q-modularity sense \cite{NewmanG04}.  Rewiring was a non-local
effect; new neighbors were chosen at random amongst all non-infected nodes in
the network rather than introducing a local structure.  It has been shown for
static networks that community structure affects disease dynamics
\cite{Grabowski2004,Huang2007}, and we expect an impact in adaptive networks
as well.

In the current work, an ideal setting was proposed where non-infected nodes were assumed to behave rationally and have perfect knowledge
of the disease status of their current neighbors and potential new neighbors.
It would be of interest to consider a situation in which not
all contagious individuals appear ill or know that they are contagious, as
might be the case for a sexually transmitted disease possessing
  asymptomatic individuals..  This effect might be
modeled by simultaneous spreading through the network of both the disease and
information about the disease. If the current model is extended with
information and community structure, social dynamics could be extrapolated to
improve contact tracing and epidemic control in organized populations with local structure.

\begin{acknowledgements}

This work was supported by the Office of Naval Research, Center for Army
Analysis, and Armed Forces Medical Intelligence Center.  The authors wish to
thank Thilo Gross and Bernd Blasius for helpful discussions.

\end{acknowledgements}

\bibliographystyle{apsrev}

\end{document}